\documentclass[9pt, conference]{IEEEtran}
\IEEEoverridecommandlockouts

\usepackage[T1]{fontenc}
\usepackage{cite}
\usepackage{amsmath,amssymb,amsfonts}
\usepackage{algorithmic}
\usepackage{graphicx}
\usepackage{textcomp}
\usepackage[dvipsnames]{xcolor}
\usepackage{moresize}

\usepackage{float}
\usepackage{lipsum}
\usepackage{listings}
\usepackage{threeparttable}
\usepackage{multirow}
\usepackage{multicol}
\usepackage{pifont}
\usepackage{booktabs}
\usepackage{moresize}

\definecolor{codegreen}{rgb}{0,0.4,0}
\definecolor{codegray}{rgb}{0.5,0.5,0.5}
\definecolor{codepurple}{rgb}{0,0,0}
\definecolor{backcolour}{rgb}{1,1,0.90}

\newcommand\notsotiny{\@setfontsize\notsotiny\@vipt\@viipt}

\lstdefinestyle{mystyle}{
    backgroundcolor=\color{backcolour},   
    commentstyle=\color{codegreen},
    keywordstyle=\color{blue},
    numberstyle=\tiny\color{codegray},
    stringstyle=\color{codepurple},
    basicstyle=\scriptsize,
    breakatwhitespace=false,         
    breaklines=true,                 
    captionpos=b,                    
    keepspaces=true, 
    numbers=left,
    numbersep=5pt,                  
    showspaces=false,                
    showstringspaces=false,
    showtabs=true,                  
    tabsize=2
}

\newfloat{lstfloat}{b}{lop}
\floatname{lstfloat}{Listing}

\def\BibTeX{{\rm B\kern-.05em{\sc i\kern-.025em b}\kern-.08em
    T\kern-.1667em\lower.7ex\hbox{E}\kern-.125emX}}
\begin{document}

\title{GAVINA: flexible aggressive undervolting for bit-serial\\mixed-precision DNN acceleration\vspace{0cm}}

\author{
    \IEEEauthorblockN{Jordi Fornt\IEEEauthorrefmark{1}\IEEEauthorrefmark{2}, Pau Fontova-Musté\IEEEauthorrefmark{1}, Adrian Gras\IEEEauthorrefmark{1}\IEEEauthorrefmark{2}, Omar Lahyani\IEEEauthorrefmark{1}, Martí Caro\IEEEauthorrefmark{1}\IEEEauthorrefmark{2} \\ Jaume Abella\IEEEauthorrefmark{1}, Francesc Moll\IEEEauthorrefmark{2}\IEEEauthorrefmark{1}, Josep Altet\IEEEauthorrefmark{2}} \vspace{2 mm}
    \IEEEauthorblockA{\IEEEauthorrefmark{1}Barcelona Supercomputing Center (BSC)
    - \{jordi.fornt, pfontova, agraslop, omar.lahyani, mcaroroc, jaume.abella\}@bsc.es}
    \IEEEauthorblockA{\IEEEauthorrefmark{2}Universitat Politècnica de Catalunya (UPC)
    - \{francesc.moll, josep.altet\}@upc.edu}
}

\maketitle

\begin{abstract}
Voltage overscaling, or undervolting, is an enticing approximate technique in the context of energy-efficient Deep Neural Network (DNN) acceleration, given the quadratic relationship between power and voltage. Nevertheless, its very high error rate has thwarted its general adoption. Moreover, recent undervolting accelerators rely on 8-bit arithmetic and cannot compete with state-of-the-art low-precision (<8b) architectures. To overcome these issues, we propose a new technique called Guarded Aggressive underVolting (GAV), which combines the ideas of undervolting and bit-serial computation to create a flexible approximation method based on aggressively lowering the supply voltage on a select number of least significant bit combinations. Based on this idea, we implement GAVINA (GAV mIxed-precisioN Accelerator), a novel architecture that supports arbitrary mixed precision and flexible undervolting, with an energy efficiency of up to 89~TOP/sW in its most aggressive configuration. By developing an error model of GAVINA, we show that GAV can achieve an energy efficiency boost of 20\% via undervolting, with negligible accuracy degradation on ResNet-18.
\end{abstract}



\section{Introduction} \label{sec:intro}

The quest for high energy efficiency is one of the main challenges of today's Deep Neural Network (DNN) hardware accelerators. Among the many techniques proposed to lower energy consumption, voltage scaling is a common practice. Lowering the supply voltage results in inflated critical path delays, so the clock frequency is typically reduced as the voltage decreases.

For applications not requiring exact computation, such as DNN inference, one may lower the supply voltage while keeping the frequency constant, allowing timing violations (and therefore computational errors) to occur. This is known as voltage overscaling (VOS), voltage underscaling, or \textit{undervolting}. Accelerators featuring undervolting can be classified into two broad classes \cite{Thundervolt_1}: Timing Error Detection and Recovery (TED) and Timing Error Propagation (TEP) accelerators. TED schemes propose circuits to detect timing errors and discard the associated results by dropping the values (i.e. setting them to zero) \cite{Thundervolt_1, Shin_2019, Thundervolt_2, Huang_2021}. On the other hand, TEP accelerators allow timing errors to propagate, and use different techniques to mitigate their impact. Some works limit the approximation by only applying undervolting to the LSBs of multipliers and/or memories, effectively protecting the MSBs \cite{configurable_multiplier, partial_uv_multiplier, xndvdla}. Recently, \cite{xtpu} proposed a neuron-wise voltage assignment by adding voltage selection bits to the DNN weights and dynamically modulating the supply voltage to retain high model accuracy.

Despite achieving a significant energy efficiency boost via undervolting, all currently proposed accelerators leveraging this technique rely on fixed-precision 8-bit MAC units. Low-precision quantization (i.e. less than 8 bits) has proven to be one of the most successful strategies for energy-efficient DNN acceleration \cite{petaopsw, EnergyCostModeling}, and the literature shows that low- and mixed-precision accelerators easily outperform their 8-bit counterparts in terms of energy efficiency (see~Fig.~\ref{fig:sota_accels}). In the current state-of-the-art, the energy-accuracy tradeoff provided by quantization vastly overshadows the benefits of undervolting. Hence, accelerators that exploit undervolting without considering low- and mixed-precision cannot compete with exact state-of-the-art low-precision architectures.

In this work, we aim to enable undervolting as a viable source of approximation for state-of-the-art DNN accelerators by combining the ideas of undervolting, bit-serial computing, and mixed-precision operation. Bit-serial computing is precision-reconfigurable, which allows to fully exploit the benefits of mixed-precision \cite{mix_precision_0, QuantizationPruning}, and is also highly compatible with emerging state-of-the-art architectures such as Compute In-Memory (CIM) \cite{Lee24, Yi24}. Moreover, bit-serial accelerators have a remarkable synergy with undervolting: the MSBs and LSBs of the computation can be processed at different times, so the supply voltage can be modulated depending on the bit significance being computed. Based on this idea, our work reports the following contributions:

\begin{figure}[t]
  \centering
\includegraphics[width=0.49\textwidth]{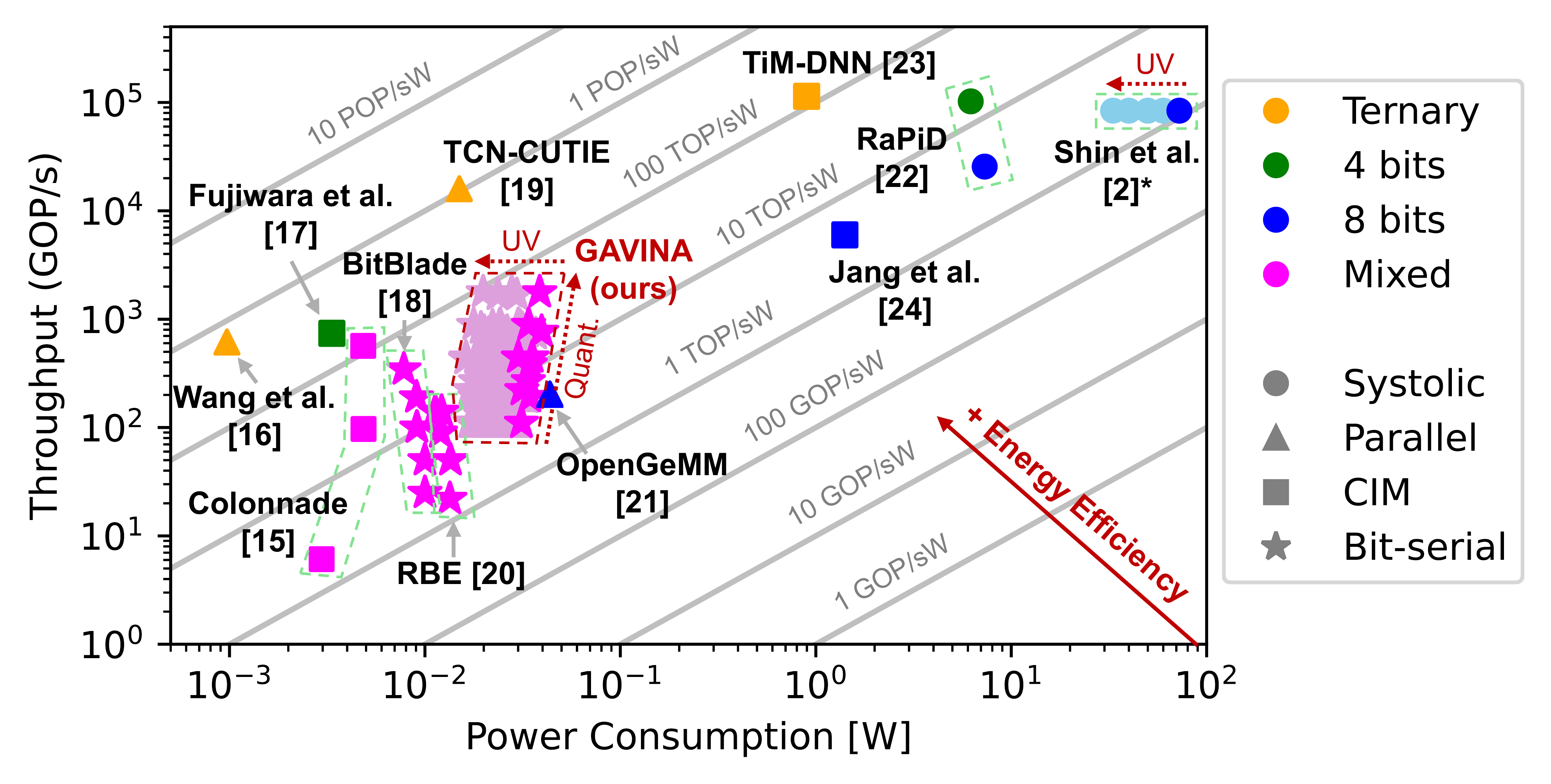}
\caption[]{Summary of digital state-of-the-art DNN accelerators \cite{colonnade, wang22, fujiwara22, bitblade, cutie, marsellus, opengemm, rapid, timdnn, jang24}. UV stands for undervolting. \textsuperscript{*}\textit{Note}: results for \cite{Shin_2019} only include the MAC array. }\vspace{-0.2cm}
\label{fig:sota_accels} 
\end{figure}

\begin{itemize}
    \item We propose a novel undervolting technique called Guarded Aggressive underVolting (GAV), based on varying the voltage supply depending on the bit significance being computed.
    \item We develop GAVINA (GAV mIxed-precisioN Accelerator), a novel bit-serial DNN accelerator based on the idea of GAV.
    \item We perform a full physical design of GAVINA and use it for an in-depth evaluation of its power and error characteristics.
    \item We develop a heuristic model of GAVINA with undervolting to evaluate its approximation effect on DNN applications, overcoming the performance bottleneck of circuit simulations.
    \item We benchmark the effects of applying GAV to ResNet-18 using the CIFAR-10 dataset.
\end{itemize}

The rest of the paper is laid out as follows. Sec.~\ref{sec:gav} explains the GAV technique in detail. Sec.~\ref{sec:arch} covers the architectural details of the GAVINA accelerator. Sec.~\ref{sec:eval} reports our evaluation results, and Sec.~\ref{sec:sota} compares our work with other state-of-the-art accelerators. Finally, Sec.~\ref{sec:conclusions} presents our conclusions.




\section{Guarded Aggressive underVolting (GAV)} \label{sec:gav}

One of the main challenges of undervolting is that bits with high significance typically have large error rates, since they have longer combinational delays due to carry chain structures. The main idea of our Guarded Aggressive underVolting (GAV) is to modulate the degree of undervolting depending on the bit significance being processed, by leveraging the properties of bit-serial computing. Listing~\ref{lst:bitserial} shows the pseudocode of a simple bit-serial general matrix-matrix multiplication (GEMM) of two matrices $\textbf{A}$ and $\textbf{B}$, of sizes $[C,L]$ and $[K,C]$ respectively, where we apply the principle of GAV.

\begin{lstlisting}[float, caption={Bit-serial GEMM loop using GAV.}, captionpos=b, language=Python, label=lst:bitserial, basicstyle=\ssmall, mathescape=true]
for ba in range(A_bits):
  for bb in range(B_bits):
    $\textcolor{orange}{\textbf{\texttt{set\_vsupply}}}$(ba,bb) # Supply voltage depends on bit significance
    for l in range(L):
      for k in range(K):
        for c in range(C):
          sign = (-1)*((ba==A_bits-1)^(bb==B_bits-1))
          $\textbf{P}$[k][l] += sign*($\textbf{A}$[c][l][ba] & $\textbf{B}$[k][c][bb]) << (ba+bb)
\end{lstlisting}

By modifying the supply voltage depending on the bit significance indices $ba$ and $bb$, the degree of approximation caused by undervolting can be modulated. Unlike in previous works, where the only option for reconfigurability is modifying the voltage of approximate power domains associated with a fixed number of bits, GAV provides three degrees of freedom which are fully reconfigurable at runtime:

\begin{enumerate}
    \item The voltage values (i.e. Dynamic Voltage Scaling or DVS).
    \item The bit significance where each voltage value is applied.
    \item The precision of both input operands (i.e. number of bits).
\end{enumerate}

\begin{figure}[t]
  \centering
\includegraphics[width=0.49\textwidth]{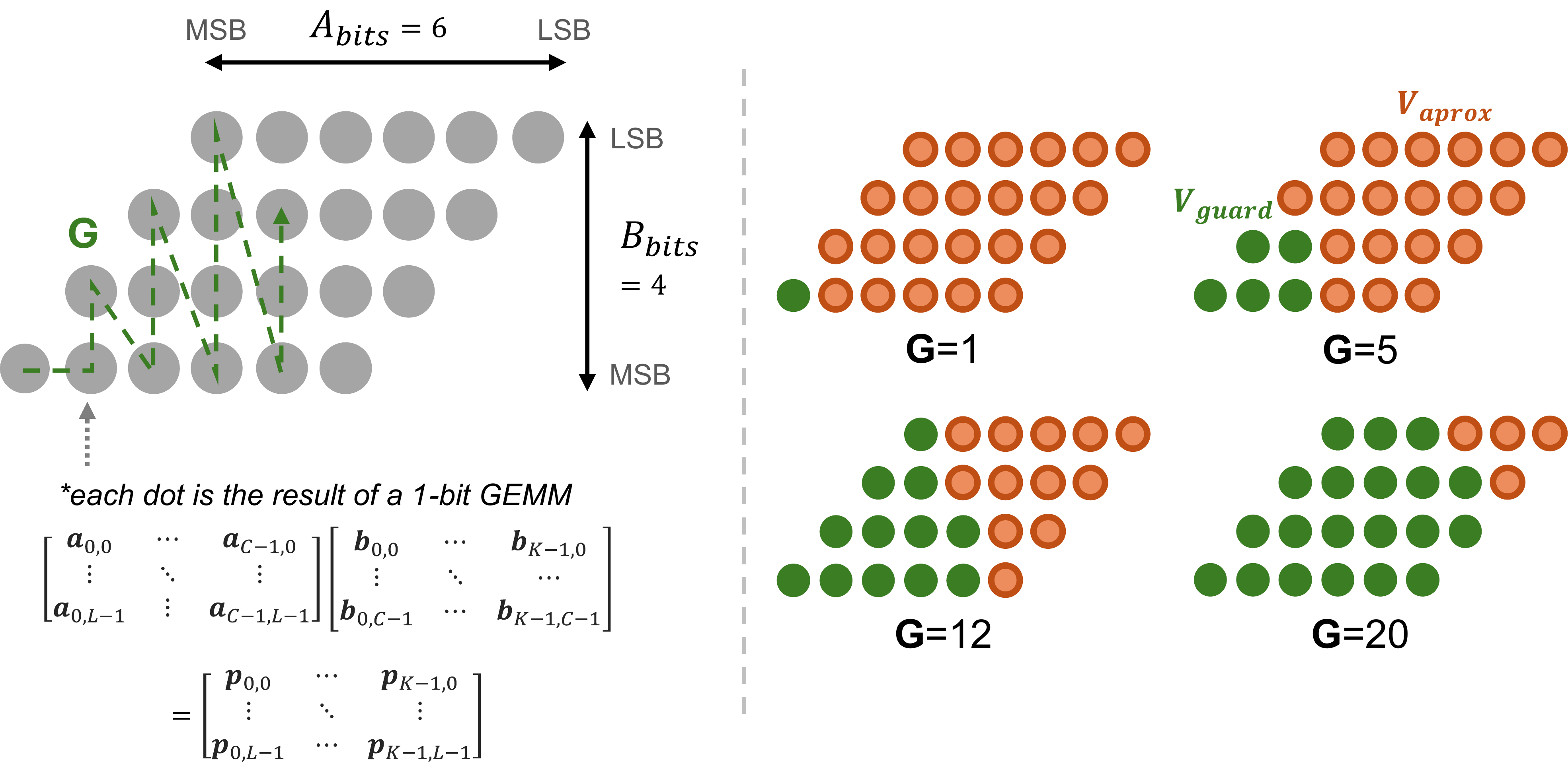}
\caption[]{GAV schedule based on a single variable G and two voltage levels.}\vspace{-0.2cm}
\label{fig:gav} 
\end{figure}

This makes GAV a very flexible approximate technique with a wide range of operating points. In this work, we focus on exploring (2) and (3), using a voltage scheduling policy that selects between two voltage levels (the \textit{guarded voltage} and the \textit{approximate voltage}), depending on the bit significance of the computation. We modulate the approximation with a single variable $G$, as depicted in Fig.~\ref{fig:gav}.

Note that this approach can be extended to more sophisticated policies with several voltage values instead of two. Nevertheless, we chose a simple configuration with only two voltage levels to focus on the main novelty introduced with GAV: dynamically modulating undervolting based on computation significance.


\section{GAV mIxed-precisioN Accelerator (GAVINA)} \label{sec:arch}

GAVINA is built around a Parallel Array of AND gates and adder trees that computes the multiplication of two 1-bit matrices of size $[C,L]$ and $[K,C]$ every clock cycle. This central compute block fully parallelizes the loops defined in lines 4-6 of Listing~\ref{lst:bitserial}. The bit precision loops (lines 1-2) are handled sequentially, allowing our accelerator to perform GEMMs with arbitrary mixed precision, and to modify the supply voltage depending on the bit significance.

\begin{figure*}[t!]
  \centering
\includegraphics[width=0.87\textwidth]{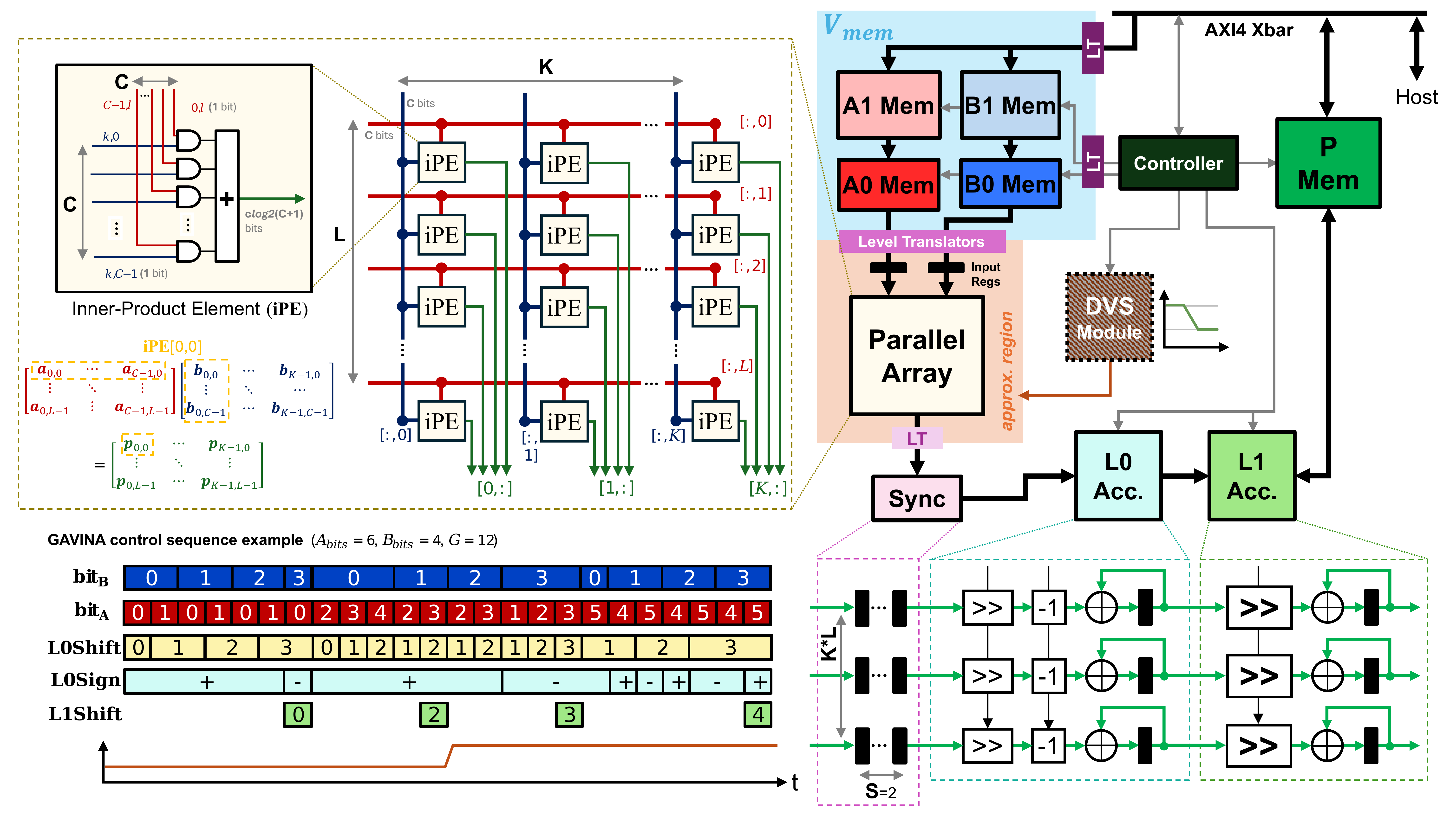}
\caption[]{GAVINA architecture diagram and an example of multi-bit integer GEMM using GAV. In the example, $bit_A$ and $bit_B$ are the control signals that index the bit significance positions of the activation and weight matrices, respectively.}
\label{fig:arch} 
\end{figure*}

As depicted in Fig.~\ref{fig:arch}, GAVINA comprises 5 logic modules plus 5 memory blocks. \textit{A1 Mem}, \textit{B1 Mem} and \textit{P Mem} hold the three matrices involved in the computation. We include an additional level of memory hierarchy for the inputs and weights (\textit{A0 Mem} and \textit{B0 Mem}) to exploit data locality during the bit-serial operation. The activation and weight matrices are stored in bit-serial format, i.e. with the operand bits not contiguous in memory. Each read operation to \textit{A0 Mem} and \textit{B0 Mem} fetches two binary matrices of size $[C,L]$ and $[K,C]$, respectively.

The Parallel Array standard cells are organized as a 2D array of Inner-Product Elements (iPEs), as shown in Fig.~\ref{fig:arch}. Every clock cycle, the Parallel Array outputs a matrix of size $[K,L]$ containing unsigned integer values that use $ceil(log_2(C+1))$ bits.

The outputs of each iPE are shifted according to their bit significance and accumulated to obtain the full integer GEMM result. Since barrel shifters are power-hungry, we split the shift and accumulate process in two stages. The Level-0 (\textit{L0 Acc.}) stage contains barrel shifters of reduced size (i.e. supporting a small range of shift positions) that are accessed every clock cycle, as well as the sign inversion and a set of registers. The Level-1 (\textit{L1 Acc.}) stage performs the final shift and accumulation operation with full-sized barrel shifters and accumulator registers, and is only accessed once after the L0 accumulation is complete.

Undervolting is implemented by defining a separate power domain for the Parallel Array and the input registers that cut the path delays between memory and logic. We denote this domain as the \textit{approximate region}, and the rest of the circuit is the \textit{protected region}. The GAV technique is applied to the approximate region by selectively lowering its supply voltage depending on the current significance being computed. By applying undervolting to the binary GEMM only, we avoid catastrophic errors related to control signals or values stored in memories.

We also define a third power domain for all the input operand memories, denoted as the \textit{memory region} ($V_{mem}$ on Fig.~\ref{fig:arch}). The purpose of this domain is not to apply undervolting, but rather to use a constant lower voltage (without allowing timing violations) to reduce the power of the memories without causing potentially catastrophic errors. Since the different power domains have independent voltage supply lines, a set of Level Translators (i.e. level shifters) is added between each pair of domains during logic synthesis.

Applying undervolting to the approximate region results in frequent timing violations on the first flip-flops that sample the approximate results. These violations may result in the flip-flops entering a metastable state. When no synchronization is used, the Mean Time Before Failure (MTBF)~\cite{metastability} of a single flip-flop using 12~nm technology, assuming an operating frequency of 100~MHz, is in the order of 1~$\mu$s. Hence, metastability in systems that allow timing error propagation is guaranteed. If left unattended, a metastable flop output can propagate through the design and cause additional unexpected timing violations in the protected region, potentially causing catastrophic errors. To prevent this, we include a synchronization stage in the protected power domain, featuring a set of 2-stage synchronizer standard cells that ensure an MTBF of over $10^{70}$ years with the operating conditions described in Sec.~\ref{sec:eval}.

The Controller module depicted in Fig.~\ref{fig:arch} is responsible for generating the memory access sequences during computation, as well as properly setting the L0 and L1 shift and sign values to compute multi-bit GEMM operations with our bit-serial scheme. Moreover, it also controls the supply voltage of the approximate region according to the GAV schedule defined in Sec.~\ref{sec:gav}. The bottom left of Fig.\ref{fig:arch} depicts an example of the GAVINA control sequence for a specific bit precision and GAV configuration.

The power supply of the approximate domain is driven by a DVS module with two output modes controlled by GAV: a guarded mode at $V_{guard}$ and an approximate mode at $V_{aprox}$. Note that this approach can be extended to any number of discrete voltage levels; we choose to consider two for simplicity. The design and implementation of the DVS module is out of the scope of this paper. Fast DVS converter design is an active field of research, and recent works have reported transition slopes over 100~mV/ns \cite{fast_dvs_converter}, suggesting that the transition between power modes under the parameters described in Sec.~\ref{sec:eval:physical} can be performed in much less than 1 clock cycle.




Tiling is used to compute GEMM operations with matrices larger than $[C,L]\times[K,C]$. All memories are double-buffered to avoid stalls during context switches, so different contexts can be concatenated without a performance penalty. By leveraging bit-serial computation, GAVINA can multiply two integer matrices $\textbf{A}[C,L]$ and $\textbf{B}[K,C]$ of arbitrary precision in $A_{bits} \cdot B_{bits}$ cycles. Hence, its maximum throughput is $\frac{L\cdot C\cdot K}{A_{bits}\cdot B_{bits}}$ MACs/cycle. By applying our novel GAV technique to this architecture, we reduce the power consumption of the accelerator while maintaining the same throughput, at the cost of sacrificing accuracy due to functional errors caused by undervolting.





\section{Evaluation} \label{sec:eval}

Our experimental methodology is laid out as follows. First, we perform a full physical design implementation of GAVINA in 12~nm technology (Sec.~\ref{sec:eval:physical}). Second, using the post-layout netlist and its associated delays, we perform a set of Gate-Level Simulations (GLS) to analyze the errors caused by undervolting (Sec.~\ref{sec:eval:efficiency_error}). Third, we develop an error model of GAVINA with undervolting, since using GLS for any non-trivial application is intractable (Sec.~\ref{sec:eval:model}). Finally, we use our undervolting model to benchmark the novel GAV technique on a DNN application (Sec.~\ref{sec:eval:dnn}).

\subsection{Physical Design} \label{sec:eval:physical}

\begin{table}[b!]
\vspace{-3mm}
\caption{GAVINA specifications (post-layout)}\vspace{-0.2cm}
\begin{center}
\begin{tabular}{|l|c|c|}
\hline
Technology& \multicolumn{2}{c|}{GF12LPPLUS (12 nm)} \\
\hline
Chip Area& \multicolumn{2}{c|}{1.60 mm x 2.10 mm} \\
\hline
Parallel Array Size ($C$x$L$x$K$)& \multicolumn{2}{c|}{73728 (576x8x16)} \\
\hline
Total memory& \multicolumn{2}{c|}{74 kB (x2)} \\
\hline
Clock Period / Frequency & \multicolumn{2}{c|} {20.0 ns / 50 MHz} \\
\hline
Max. Throughput ($a2w2$)& \multicolumn{2}{c|}{1.84 TOP/s} \\
\hline
$V_{mem}$& \multicolumn{2}{c|}{0.40 V} \\
\hline
$V_{guard}$   $\vert$   $V_{aprox}$ & \hspace{0.09cm} 0.55 V \hspace{0.09cm} & 0.35 V\\
\hline
Avg. Power @ Peak TOP/s & 38.67 mW  & 19.86 mW \\
\hline
\end{tabular} \vspace{-0.4cm}
\label{tab_specs}
\end{center}
\end{table}

\begin{figure}[b!]
  \centering
\includegraphics[width=0.48\textwidth]{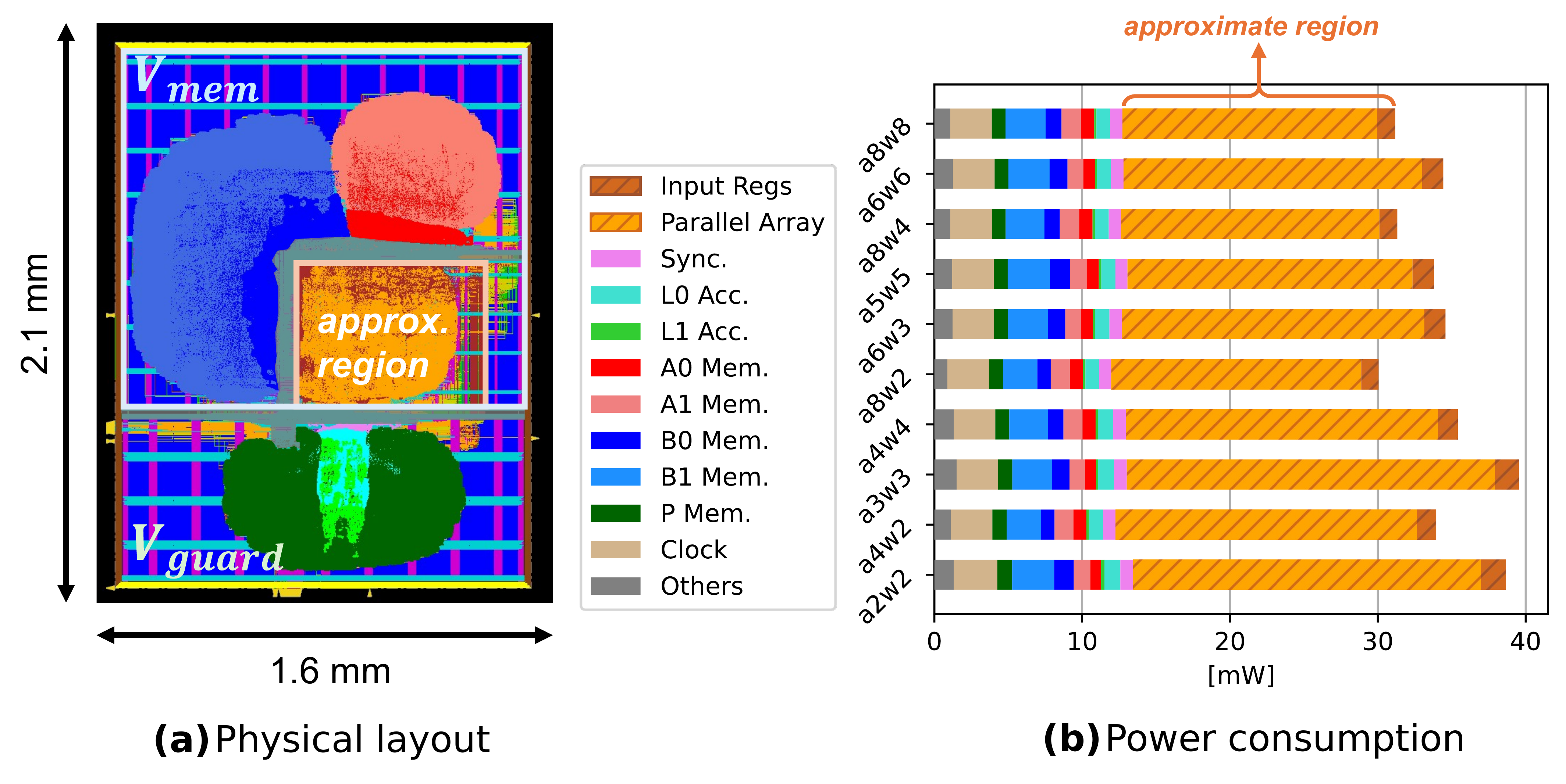}
\caption[]{Post-layout floorplan (a) and power distribution of GAVINA for different precision configurations (b), without undervolting (using $V_{guard}$).}\vspace{-0.2cm}
\label{fig:floorplan} 
\end{figure}

We use the architectural parameters $[C,L,K]$ = $[576,8,16]$ to perform the physical design of GAVINA. This particular shape is motivated by two facts: first, our architecture benefits from having a large number of input channels ($C$), as this dimension is reduced at the Parallel Array level, so the subsequent modules only scale by $O(log(C))$. Second, considering that we use a Convolutional Neural Network (CNN) as a benchmark in Sec.~\ref{sec:eval:dnn}, it is convenient to have the dimension $C$ be a multiple of 9, since most CNNs typically use kernels of size $3\times3$ in their inner layers.

All memory blocks are implemented using Standard Cell Memories (SCM) based on latches. Even though SCMs require more area than SRAMs or register files, they have a much smaller power footprint and can take full advantage of clock gating \cite{cutie}. Our experiments with GAVINA show that using SCMs instead of SRAMs results in a power reduction of about $\times4$ and an area increase of $\times2$.

During the synthesis and place-and-route of the accelerator, we use the standard cell libraries characterized at $V_{guard}$ in the approximate region. After the backend process is finished, we characterize the netlist delays under $V_{aprox}$. This way, we prevent the EDA tools from trying to meet timing in the undervolting mode. Conversely, in the memory region, we consider $V_{mem}=0.40 V$ during synthesis and backend too, since we do not apply undervolting in that domain. We perform our own characterization of the standard cell library to obtain the libraries at the specific voltages used in the accelerator. The standard cell characterization, synthesis, place-and-route and gate-level simulations are performed with Cadence EDA tools.

Table~\ref{tab_specs} summarizes the specifications of the implemented version of GAVINA, and Fig.~\ref{fig:floorplan} depicts the floorplan of the accelerator and a breakdown of its power consumption for several precision configurations. Note that, in subsequent sections, we use the shorthand notation $aXwY$ to denote a precision of $X$ and $Y$ bits for the activations weights, respectively.


\subsection{Error and Power Consumption} \label{sec:eval:efficiency_error}

\begin{figure}[t]
  \centering
\includegraphics[width=0.45\textwidth]{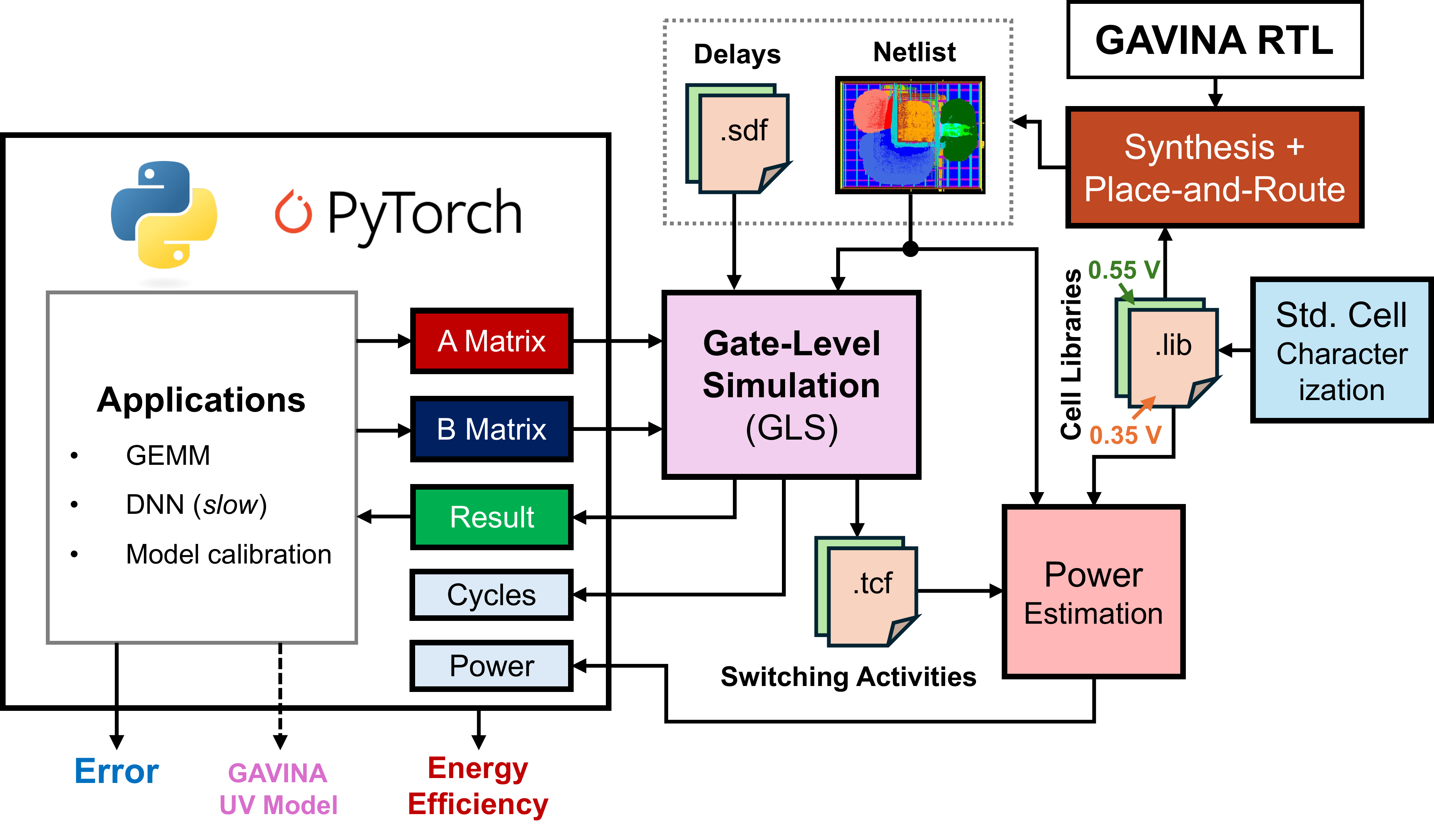}
\caption[]{Summary of our experimental methodology for error and energy efficiency estimation using GLS.}
\label{fig:methodology} 
\end{figure}

We characterize the error incurred by the accelerator when using GAV, as well as its power consumption, by performing GLS of the circuit computing a set of matrix-matrix multiplications. To consider the effects of only undervolting part of the computation, as defined by GAV, we split the GEMM into its exact and approximate components, and run two GLSs, each one with the delay file of its corresponding mode. Fig.~\ref{fig:methodology} provides a summary of our experimental setup.

The input matrices used for error analysis are random matrices of size $[4608,64]$ and $[64,4608]$. The values are generated using a probability distribution that forces both the inner-products computed by the GEMM and one of the input operands to follow an approximately uniform distribution. This allows us to observe the full dynamic range that can be obtained by a GEMM. We use uniform symmetric quantization \cite{quantization_1} to quantize the random matrices.


As a metric to quantify the error, we use the variance of the normalized error distance ($VAR_{NED}$), as previous works have reported that it is a good indicator of accuracy degradation in DNNs \cite{var_ned}. We define $VAR_{NED}$ as:

\begin{equation}
VAR_{NED} = \frac{1}{N} \sum_{n=1}^{N} \left( NED_i - \frac{1}{N} \sum_{n=1}^{N} NED_i\right) ^2
\end{equation}

Where $NED_i$ is the normalized error distance of a measurement, defined as $NED_i = \frac{E_i-A_i}{E_{max}}$, $E_i$ and $A_i$ are the exact and approximate values, and $E_{max}$ is the maximum absolute exact value. 

\begin{figure}[t]
  \centering
\includegraphics[width=0.49\textwidth]{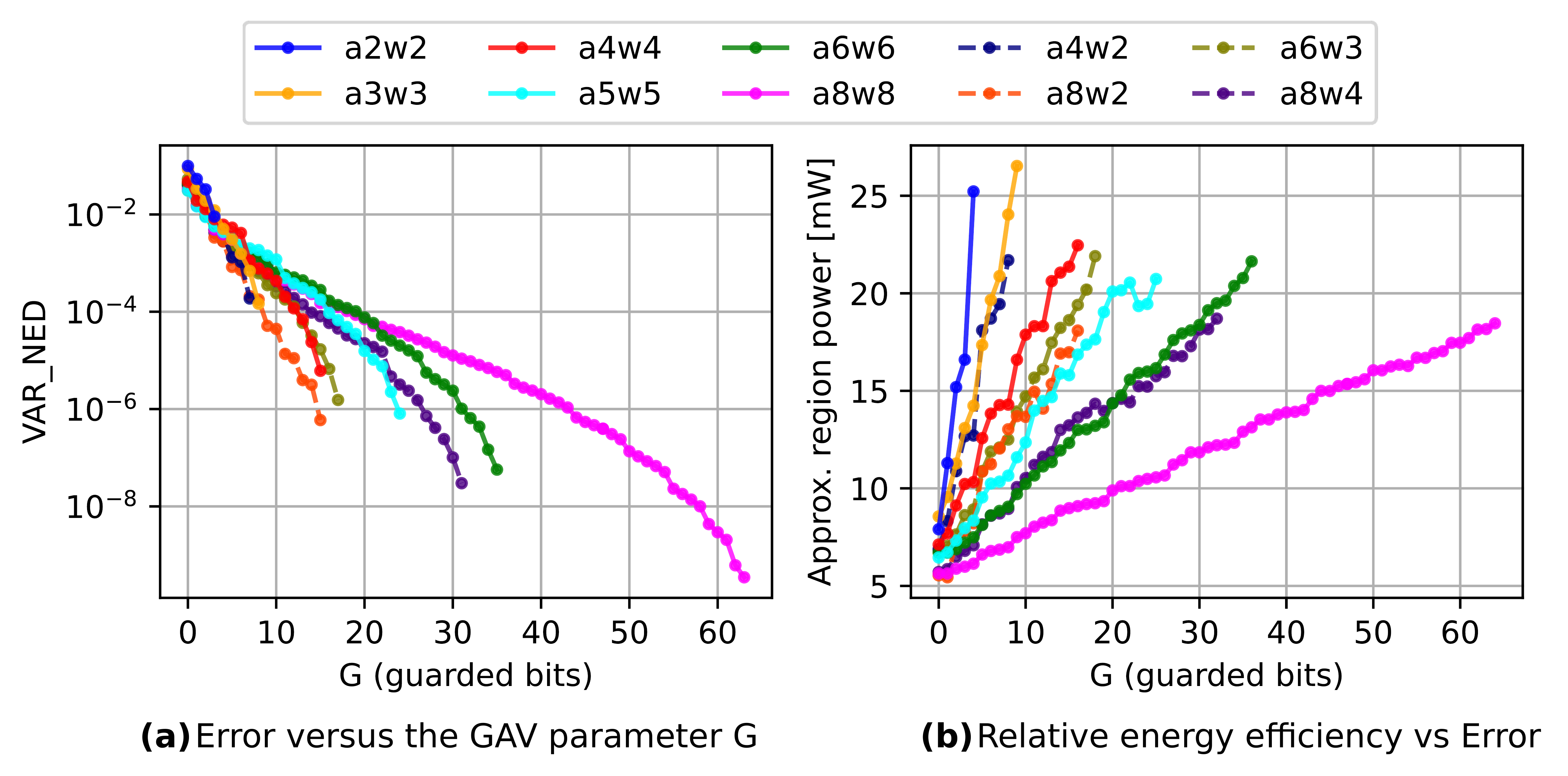}
\caption[]{$VAR_{NED}$ for different precisions and $G$ values (a). Error versus approximate region power consumption (b).}\vspace{-0.2cm}
\label{fig:error_energy_tradeoff} 
\end{figure}

As depicted in Fig.~\ref{fig:error_energy_tradeoff}a, the error caused by undervolting decreases exponentially with $G$ on all precisions, as more bits are protected. Higher precisions have more granularity, since there are more significance positions to choose from. In terms of power consumption, the approximate region power is reduced by up to $\times 3.5$ in the most aggressive configurations (see Fig.~\ref{fig:error_energy_tradeoff}b). At the accelerator level, this translates into an energy efficiency boost of up to $\times 1.95$, as other elements in the system (especially the memories) end up dominating when the main compute power is reduced.



To benchmark how the error characteristics of GAV affect the accuracy of a DNN, it is technically possible to use the experimental methodology described in Fig.~\ref{fig:methodology} by replacing the GEMMs used in the network with our setup. However, due to the complexity of GLS with delay information, this results in intractable runtimes. For example, a single forward pass of ResNet-18 with one CIFAR-10 image using 4-bit precision takes about 2 hours to process. Following this method, one evaluation of the validation dataset would take years. To overcome this, we developed a software model of the GAVINA architecture with undervolting, detailed in Sec.~\ref{sec:eval:model}.

\subsection{The GAVINA Undervolting Model} \label{sec:eval:model}

Other works have constructed models to estimate the error incurred by undervolting. The authors of \cite{xndvdla} define a set of look-up tables based on the GLS of 8-bit multiplier units featuring undervolting, resulting in tables with $2^8 \times 2^8 = 65$k entries. Unfortunately, this approach cannot be applied to GAVINA, since our undervolted circuit (the Parallel Array, see Fig.~\ref{fig:arch}) has much wider inputs that would make such look-up tables explode in complexity, requiring at least $10^{346}$ entries with the settings described in Sec.~\ref{sec:eval:physical}.

We solve this issue by defining a heuristic model based on the Parallel Array outputs rather than its inputs. By analyzing the effects of undervolting during GLS, we define a set of error probability functions that fit the circuit's behavior with the exact expected value as an input. To simplify the model, we assume that all iPEs of the Parallel Array follow the same error distribution.

\begin{figure}[t]
  \centering
\includegraphics[width=0.48\textwidth]{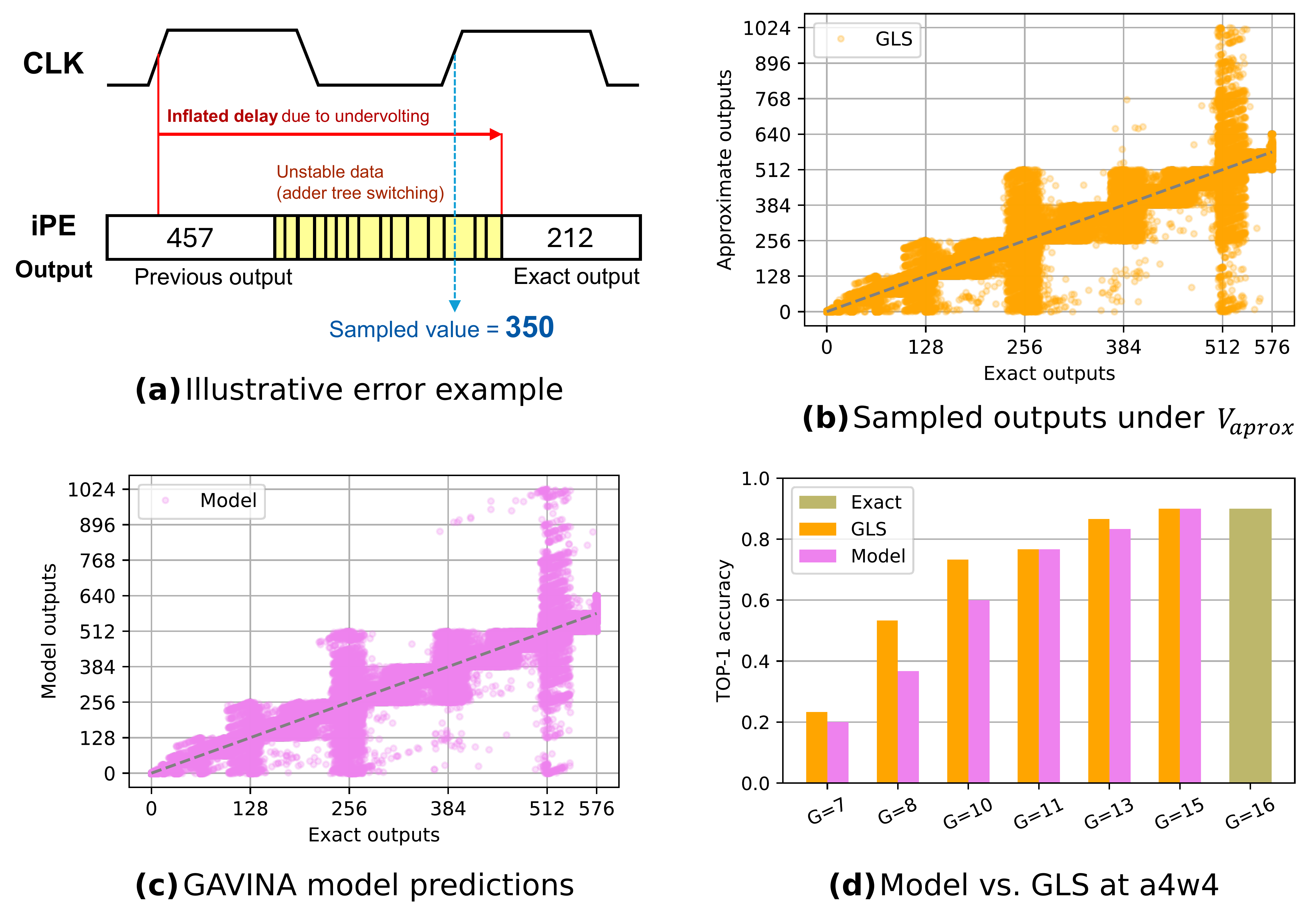}
\caption[]{Depiction of undervolting errors (a). Sync. outputs under the effect of undervolting, from GLS (b). Predicted outputs from the model (c). Comparison between model and GLS on 30 CIFAR-10 images (d).}
\label{fig:error_illustration} 
\end{figure}

Fig.~\ref{fig:error_illustration}a illustrates a functional error during undervolting. Due to the increased path delays caused by reducing the voltage supply, the iPEs of the Parallel Array take longer to produce stable outputs, and the Sync module (see Sec.~\ref{sec:arch}) samples an unstable signal. As shown in Fig.~\ref{fig:error_illustration}b, this causes the partial products to have erroneous values with a particular error distribution. By analyzing these results, we made several experimental observations that inform our heuristic model:

\begin{itemize}
    \item \textbf{Bit dependency.} Some bits of the iPE output are much more likely to have errors than others, due to the MSBs having longer carry chains and therefore longer delays.
    \item \textbf{Exact output dependency.} The error distribution depends on the exact value obtained without undervolting. In particular, some locations near power-of-two values have larger error rates.
    \item \textbf{Previous value dependency.} The previous output value influences the probability of an erroneous bit, since it determines how the iPE sum bits switch to the next output, and how long it takes for the signal to stabilize.
    \item \textbf{Neighboring bit dependency.} If a particular bit is erroneous, the probability of errors in neighboring bits is higher.
\end{itemize}

From these observations, we define a look-up table with four indexing dimensions, each corresponding to one of the dependencies above. The values in the table represent the probability of a particular bit of the iPE output being flipped under a certain set of conditions. To model the effect of neighbors, we consider each bit to have a conditional dependency on the bits with more significance: e.g., in an iPE with a 10-bit output $[b_0,b_1,...,b_9]$, the error probability of $b_7$ depends on whether or not $[b_8,b_9]$ have had errors. The MSB ($b_9$) is the only bit that does not have a conditional dependency.

To improve performance, we reduce the size of the look-up tables by considering some simplifications. First, we only consider the $n_{nei}$ higher significance neighbors, limiting the possible conditions to $N_{cond}$ = $2^{n_{nei}}$ (e.g. if $n_{nei}$=$2$, the error probability of $b_3$ depends on $[b_4,b_5]$, with a total of $2^2$=$4$ possible conditions depending on the sampled outcome for $[b_4,b_5]$). Second, we store the model as a ragged list of tables with different sizes, since some bits have fewer neighbors with higher significance (e.g. the MSB contemplates no neighbors). Third, we simplify the previous value index by splitting the range of possible values into a fixed number of bins ($p_{bins}$).

The probability tables of the GAVINA model are calibrated by filling the look-up tables with empirical error frequencies obtained from running GLS. Once calibrated, the model uses the tables as a probability distribution to randomly sample errors. The model used for evaluation has the following parameters: $[n_{nei},p_{bins}]$=$[2,16]$.


\lstdefinelanguage{gavinatorch}{%
  language     = Python,
  morekeywords = {load, zeros, rand},
}

\begin{lstlisting}[float=t, caption={GAVINA model pseudocode (simplified).}, captionpos=b, language=gavinatorch, label=lst:model, basicstyle=\tiny, numbers=right, mathescape=true]
def $\textbf{\texttt{gavina\_model}}$(A, B, cal_file):
    a_seq = $\textcolor{magenta}{\texttt{reshape\_n\_slice\_a}}$(A)    # shape: [C,L,1,seqlen], dtype=bool
    b_seq = $\textcolor{magenta}{\texttt{reshape\_n\_slice\_b}}$(B)    # shape: [C,1,K,seqlen], dtype=bool

    $\textbf{\texttt{P\_TABLES}}$ = load(cal_file)  # indices: [S_BITS,C+1,pbins,Ncond]

    # Compute exact outputs and previous values
    exact_output = (a_seq & b_seq).sum(axis=0)  # shape: [L,K,seqlen]
    prev_output = exact_output[:,:,1:]
    prev_bins = $\textcolor{magenta}{\texttt{create\_bins}}$(prev_output)

    # Initialize error structures
    bit_errs = torch.zeros()    # shape: [S_BITS,L,K,seqlen]
    err_mask = torch.zeros()    # shape: [L,K,seqlen]

    # Iterate from sum MSB to LSB
    for bit in S_BITS-1..0:   
        nei_cond = $\textcolor{magenta}{\texttt{get\_conditions}}$(bit_errs[bit+1:bit+1+n_nei])
        error_prob = $\textbf{\texttt{P\_TABLES}}$[bit][exact_output,prev_bins,nei_cond]

        rand_seed = torch.rand()    # shape: [L,K,seqlen]
        bit_errs[bit] = rand_seed < error_prob
        err_mask = err_mask | (bit_errs[bit] << bit)

    # Apply sampled errors by flipping exact bits
    approx_output = exact_output^err_mask
    return $\textcolor{magenta}{\texttt{reshape\_c}}$(approx_output)
\end{lstlisting}

Listing~\ref{lst:model} shows a simplified pseudo-code of the model. The first step is to slice the input matrices into their bit components and reshape them to follow the same sequence as in GAVINA, since the order of operations is important due to the previous value dependency. The exact outputs and previous values are computed by performing the exact iPE operation (AND and sum). Starting from the MSB, the model gets the error probability of the iPE outputs by indexing the look-up tables with the current conditions. Then, it samples the predicted errors according to the inferred probabilities, and uses these predictions to flip some bits of the exact outputs.

The GAVINA undervolting model is written using the PyTorch library, and can be executed on GPU devices. As shown in Fig.~\ref{fig:error_illustration}c, the model accurately represents the error of the iPEs with undervolting. When running the same error characterization as in Sec.~\ref{sec:eval:efficiency_error} using the model, the $VAR_{NED}$ metrics are within an 8\% range of the GLS results, on average. We verify the applicability of the model to a real DNN target by running a small set of CIFAR-10 images through ResNet-18 using the GLS setup from Fig.~\ref{fig:methodology}. As shown in Fig.~\ref{fig:error_illustration}d, the accuracy results are very similar, with our model being slightly pessimistic. The GAVINA model using $a4w4$ only takes about 0.2~s per image, an acceleration of about $\times3.6\cdot10^4$ with respect to GLS.



\subsection{DNN Benchmarking} \label{sec:eval:dnn}

To assess the viability of applying GAV in DNNs, we perform several experiments with the ResNet-18 network \cite{resnet} and the CIFAR-10 image classification task. We use the error model described in Sec.~\ref{sec:eval:model} to estimate the network degradation with undervolting. The power consumption of GAVINA is estimated following the methodology explained in Sec.~\ref{sec:eval:efficiency_error}, considering a small subset of activations and weights to generate switching activities with GLS. We use the Brevitas library \cite{brevitas} to quantize and retrain the network under different precisions. To maintain as much accuracy as possible in low precisions, we progressively retrain the model from high to low precision: e.g. the $a2w2$ model is retrained from the $a3w3$ weights, which were retrained from $a4w4$, and so on.

Applying the same degree of undervolting to the whole network is naive, as some layers are more sensitive to perturbations than others. We take inspiration from \cite{xtpu} to develop an optimization algorithm that finds the optimal per-layer allocation of $G$ based on an integer linear programming (ILP) approach. Instead of minimizing an estimate of energy consumption, as in \cite{xtpu}, we choose to minimize the perturbation of the network outputs to retain as much accuracy as possible. To measure this perturbation, we apply undervolting to each layer independently for different values of $G$, and capture the mean squared error (MSE) of the network output with respect to the exact result. We constrain the problem by setting a target average $G_{tar}$ such that $weigh\_avg([G_0,...G_{L-1}])<G_{tar}$, where $L$ is the number of layers, $G_l$ is the GAV configuration assigned to layer $l$ and $weigh\_avg$ is the weighted average taking into account the number of operations on each layer.



\begin{figure}[t]
  \centering
\includegraphics[width=0.49\textwidth]{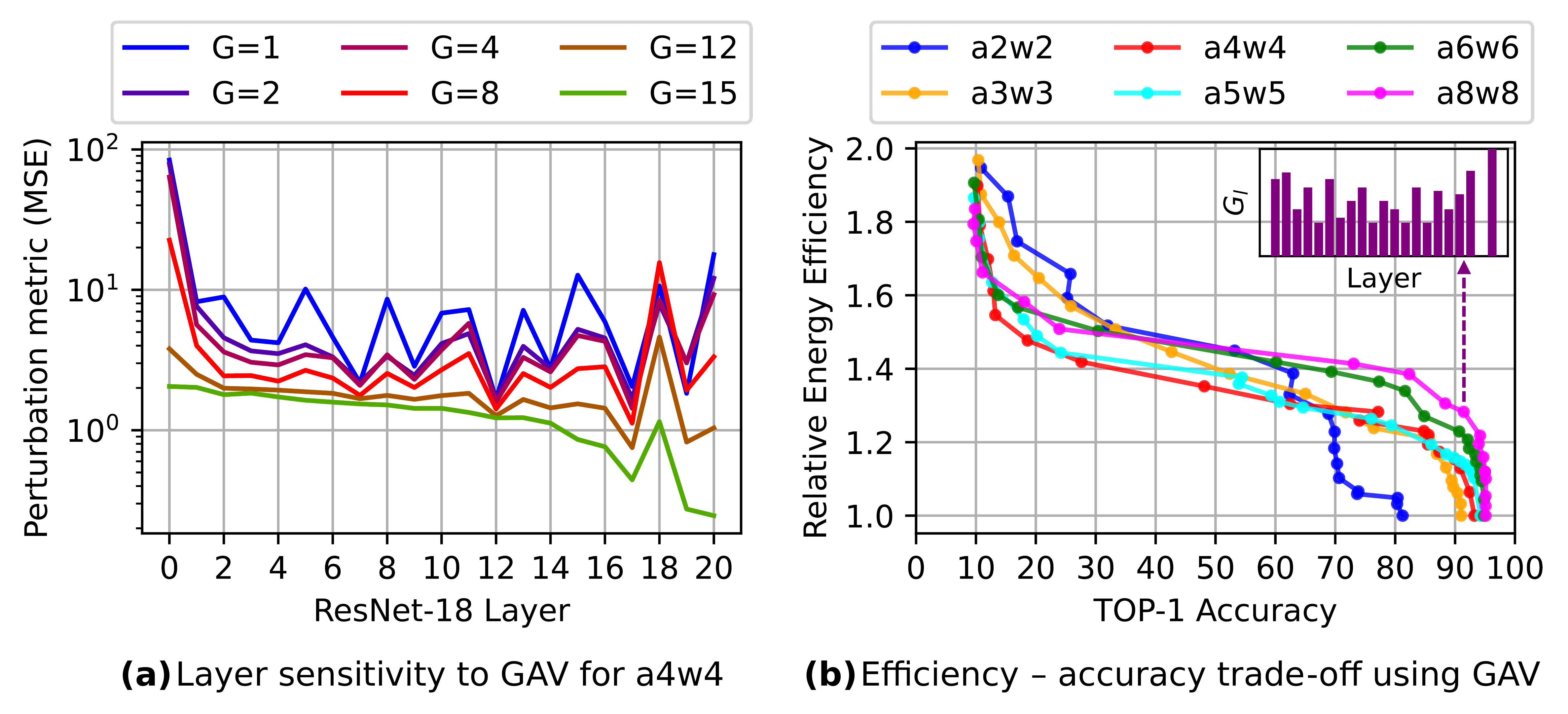}
\caption[]{Perturbation caused by GAV for different values of $G$ (a). Energy efficiency-accuracy trade-off of GAVINA using our ILP-based GAV allocation algorithm (b).}\vspace{-0.3cm}
\label{fig:dnn_results} 
\end{figure}

Fig.~\ref{fig:dnn_results}a shows the MSE metrics captured for the $a4w4$ case. As expected, GAV affects the network layers differently, with some (e.g. the input layer) being extremely sensitive to it. Our ILP approach ensures that such layers are automatically assigned larger values for $G$, while lower values are applied to the least critical layers. The energy-accuracy characteristics of different optimal GAV configurations are summarized in Fig.~\ref{fig:dnn_results}b. Low-precision configurations experience a sharper accuracy degradation since the network is already under heavy quantization noise. Higher precisions can maintain a negligible accuracy drop while gaining up to a 20\% energy efficiency boost.

\section{Comparison to state-of-the-art} \label{sec:sota}

The landscape of DNN accelerators in the recent literature is vast, as Fig.~\ref{fig:sota_accels} suggests. In this section, we compare GAVINA to a small subsample of the most similar architectures in the state-of-the-art.

Even without its undervolting features, GAVINA is a competitive accelerator in terms of energy efficiency, as summarized in Tab.~\ref{tab:related_cmp}. With its fully guarded configuration, our architecture is $\times2.08$ more energy efficient than \cite{marsellus}, a recent mixed-precision accelerator also featuring a bit-serial approach. BitBlade \cite{bitblade} is more energy efficient than GAVINA when accounting for the technology difference \cite{deepscaletool}, but it does not support fine-grained mixed-precision, and its traditional voltage scaling requires lowering its throughput significantly to increase energy efficiency. GAVINA, on the other hand, can use aggressive undervolting as an efficiency boost without affecting its throughput or fine-grained mixed-precision support.


Compared to \cite{Shin_2019}, which uses a TED approach on an 8-bit systolic architecture, our accelerator is much more versatile thanks to its mixed-precision support. On its lowest bit precision and without undervolting, GAVINA is $\times3.04$ times more efficient than \cite{Shin_2019} with its most aggressive voltage. In terms of the energy efficiency boost from using undervolting, \cite{Shin_2019} reports a maximum of $\times2.2$, but it only considers the MAC array, disregarding other power-hungry blocks like memories, which have a significant impact on the overall power (see Fig.~\ref{fig:floorplan}b). In the same conditions (only considering the main compute element), GAVINA achieves a maximum boost of $\times3.5$.

\begin{table}[t]
\centering
\caption{Comparison with other state-of-the-art accelerators.}
\label{tab:related_cmp}
\begin{threeparttable}
\resizebox{0.49\textwidth}{!}{  
    \begin{tabular}{ccccccccc}
    \toprule
    && \textbf{\begin{tabular}[c]{@{}c@{}}RBE\\ \cite{marsellus}\end{tabular}} & \textbf{\begin{tabular}[c]{@{}c@{}}BitBlade\\ \cite{bitblade}\end{tabular}} & \textbf{\begin{tabular}[c]{@{}c@{}}Shin et al.\\ \cite{Shin_2019}\end{tabular}} & \textbf{\begin{tabular}[c]{@{}c@{}}X-NVDLA\\ \cite{xndvdla}\end{tabular}} & \textbf{\begin{tabular}[c]{@{}c@{}}X-TPU\\ \cite{xtpu}\end{tabular}} & \textbf{\begin{tabular}[c]{@{}c@{}}GAVINA\\ (This Work)\end{tabular}} \\
    \midrule
    \multicolumn{2}{c}{\textbf{Technology} [nm]} & 22 & 28 & 65 & 15 & 15 & 12 \\
    \multicolumn{2}{c}{\textbf{Area} [mm$^2$]} & 2.42 & 0.71 & 214 & \textit{NA}\tnote{\textdagger} & \textit{NA}\tnote{\textdagger} & 3.36 \\
    \multicolumn{2}{c}{\textbf{Frequency} [MHz]} & 100 & 44 & 641 & \textit{NA}\tnote{\textdagger} & \textit{NA}\tnote{\textdagger} & 50 \\
    \multicolumn{2}{c}{\textbf{Implementation}} & Silicon & Silicon & Post-layout\tnote{\#} & Extrapolation\tnote{\$} & Synthesis & Post-layout \\
    \multicolumn{2}{c}{\textbf{Supply Voltage} [V]} & 0.5 & 0.6 & 1.08 - 0.73 & 0.80 - 0.40 & 0.80 - 0.50 & 0.55 - 0.35 \\
    \multicolumn{2}{c}{\textbf{Precision support}} & All 8b-2b & 8b/4b/2b & 8b & 8b & 8b & All 8b-2b \\
    \multicolumn{2}{c}{\textbf{Undervolting}} & \ding{55} & \ding{55} & \ding{51} & \ding{51} & \ding{51} & \ding{51} \\
    \midrule
    \multirow{4}{*}{\textbf{TOP/s}} & $a8w8$ & 0.022 & 0.025 & 84.0 & \textit{NA}\tnote{\textdagger} & \textit{NA}\tnote{\textdagger} & 0.111 \\
     & $a4w4$ & 0.090 & 0.100 & - & - & - & 0.443 \\
     & $a3w3$ & \textit{NA}\tnote{$\delta$} & - & - & - & - & 0.776 \\
     & $a2w2$ & 0.136 & 0.344 & - & - & - & 1.774 \\
    \cmidrule(lr){2-8} 
    \multirow{4}{*}{\textbf{TOP/sW}\tnote{$\phi$}}  & $a8w8$ & 2.91 & 5.60 & 6.91 - 15.1 & \textit{+35\%\tnote{\textdagger}} & \textit{+57\%\tnote{\textdagger}} & 3.56 - 6.52 \\
     & $a4w4$ & 10.3 & 23.5 & - & - & - & 12.52 - 23.78 \\
     & $a3w3$ & \textit{NA}\tnote{$\delta$} & - & - & - & - & 19.37 - 38.13 \\
     & $a2w2$ & 22.0 & 98.8 & - & - & - & 45.87 - 89.32 \\
     \midrule
     \multicolumn{2}{c}{\textbf{Benchmark}} & Conv. & \textit{NA}\tnote{$\delta$} & ResNet-18 & ResNet-50 & ResNet-50 & ResNet-18 \\
    \end{tabular}
}{}
\begin{tablenotes}
    \tiny
    \item[$\delta$]Not reported in \cite{bitblade} and \cite{marsellus}.
    \item[$\phi$]Scaled to 12~nm using \cite{deepscaletool}, considering a linear interpolation between 10~nm and 14~nm.
    \item[\#]Results from \cite{Shin_2019} only include a $256\times256$ MAC array (memories, FIFOs, and control are excluded).
    \item[\textdagger]Both \cite{xndvdla} and \cite{xtpu} only report relative energy savings, without disclosing absolute throughput or power results.
    \item[\$]\cite{xndvdla} does not perform synthesis of the accelerator. Energy results extrapolate measurements from previous works. \vspace{-0.4cm}
  \end{tablenotes}
\end{threeparttable}
\end{table}

\cite{xndvdla} and \cite{xtpu} apply a TEP approach like GAVINA, but neither provides absolute throughput or power results that would allow us to establish a direct energy efficiency comparison. In terms of architecture, the authors of \cite{xndvdla} modify the NVDLA architecture \cite{nvdla} by integrating a set of 8-bit multipliers with undervolting only on their LSBs. Although this idea is similar to GAV, the number of undervolted bits in \cite{xndvdla} is fixed, while GAV is dynamically reconfigurable. Moreover, GAV can support any number of voltage levels (even though we evaluated GAVINA with two, for simplicity). On the other hand, \cite{xtpu} applies undervolting to all the bits of their 8-bit multiplier units, and modulates the approximation by changing the supply voltage depending on the weight values and layer index. While GAVINA does not support fine-grained weight-wise voltage tuning, it fully supports per-layer voltage tuning, besides all the features of GAV. Regarding the effect of undervolting, GAVINA achieves a maximum boost in energy efficiency of $\times 1.96$, surpassing the largest energy savings reported by \cite{xndvdla} and \cite{xtpu}. Moreover, our accelerator also takes full advantage of low- and mixed-precision quantization: from its highest precision (8-bit) to the lowest (2-bit), GAVINA gets a $\times18$ energy efficiency boost, far surpassing any undervolting-related boost reported in the literature. To the best of our knowledge, our work is the first to combine the ideas of mixed precision, undervolting, and bit-serial computation synergistically to build a DNN accelerator with dynamically reconfigurable precision and undervolting approximation.


\section{Conclusions} \label{sec:conclusions}

We present GAVINA, an energy-efficient DNN accelerator that integrates our novel Guarded Aggressive underVolting (GAV), a flexible undervolting technique based on modulating the approximate region's voltage according to the bit significance being computed. Our accelerator is dynamically reconfigurable in terms of throughput (by changing its precision) and power (by configuring the GAV schedule), enabling a wide range of operation modes. To the best of our knowledge, our work is the first to integrate bit-serial computation and undervolting in an accelerator that exploits the synergies between the two techniques. The baseline accelerator has an energy efficiency of up to 45.87~TOP/sW in the smallest precision. By applying undervolting, energy efficiency can be boosted by up to $\times 1.96$ without affecting throughput, at the cost of sacrificing accuracy.

To benchmark the GAV technique on a meaningful DNN target and dataset, we develop a heuristic model based on empirical probabilities that can accurately emulate the error characteristics of GAVINA with undervolting. Using this model, we show that our technique can produce configurations achieving a 20\% efficiency boost with negligible accuracy degradation on ResNet-18. As part of our future work, we will focus on expanding this boost by exploiting the novel GAV technique to its full potential so that more aggressive configurations can be used with minimal accuracy penalty.


\bibliographystyle{IEEEtran}
\bibliography{sample-base}

\end{document}